\documentclass[aps,prl,reprint,showpacs]{revtex4-1}

\usepackage{graphicx}

\def\grad{\mathrm{grad}}
\def\ds{\Delta S}
\def\dz{\Delta_z}
\def\dt{\delta t}
\def\Dt{\Delta t}

\begin{document}

\title{A physical description of the adhesion and aggregation of platelets} 

\author{Bastien Chopard$^1$}\email{Bastien.Chopard@unige.ch}
\author{ Daniel Ribeiro de Sousa$^2$}
\author{Jonas Latt$^1$}\author{ Frank Dubois$^3$}\author{Catherine
  Yourassowsky$^3$} \author{Pierre Van Antwerpen$^4$}
\author{Omer Eker$^5$}
\author{Luc Vanhamme$^6$}\author{David Perez-Morga$^5$} 
\author{Guy Courbebaisse$^7$} \author{Karim Zouaoui Boudjeltia$^2$}%

 \affiliation{$^1$Computer Science Department, University of Geneva, Switzerland }%
\affiliation{ $^2$Laboratory of Experimental Medicine, Université
  Libre de Bruxelles (ULB) and CHU de Charleroi, Belgium }%
\affiliation{$^3$Microgravity Research Centre, Universit\'e libre de
  Bruxelles (ULB), Belgium}
 \affiliation{$^4$Laboratory of
  Pharmaceutical Chemistry and Analytic Platform of the Faculty of
  Pharmacy, Universit\'e libre de Bruxelles (ULB), Belgium}
\affiliation{$^5$Department of Interventional Neuroradiology, CHRU de Montpellier, France}
\affiliation{$^6$Institute
  of Molecular Biology and Medicine, Universit\'e Libre de Bruxelles
  (ULB), Belgium}
\affiliation{$^7$CREATIS, INSA Lyon, University
  of Lyon, France}

\begin{abstract}
  The early stages of clot formation in blood vessels involve
  platelets adhesion-aggregation. Although these mechanisms have been
  extensively studied, gaps in their understanding still persist. We
  have performed detailed {\it in-vitro} experiments and developed a
  numerical model to better describe and understand this
  phenomenon. Unlike previous studies, we took into account both
  activated and non-activated platelets, as well as the 3D nature of
  the aggregation process. Our investigation reveals that blood
  albumin is a major parameter limiting platelet adhesion and
  aggregation. Our results also show that the well accepted
  Zydney-Colton shear-induced diffusivity is much too low to explain
  the observed deposition rate.  Simulations are in very good
  agreement with observations and provide quantitative estimates of the
  adhesion and aggregation rates that are hard to measure
  experimentally.
\end{abstract}



\maketitle

Despite their critical importance in clinical practice and physiology,
the mechanisms by which hemodynamic conditions lead to platelets
adhesion (process by which individual platelets bind to a vessel wall)
and aggregation (process by which platelets attach to each other) are
still incompletely understood. The current understanding is that at low
shear rate ($0\to 1000~s^{-1}$) platelet aggregation is primarily
mediated by soluble fibrinogen, which physically crosslinks platelets
through engagement of integrin $\alpha$IIb$\beta$3. At progressively
higher shear rate~\cite{savage:96,savage:98} ($1000\to 10000~s^{-1}$)
aggregation becomes increasingly driven by the von Willebrand Factor
through its ability to rapidly engage glycoprotein Ib, with fibrinogen
playing a supportive role in stabilizing aggregates. However, the
parameters that limit the platelet adhesion and aggregation in space
and time have not been clearly identified~\cite{maxwell:07}. A
quantitative 3D model of adhesion and aggregation will reveal the
respective importance of the various mechanisms at play, thus offering
the possibility to control the platelets behavior in patients with
cardiovascular risk factors with a proper dosage of pharmacological
agents.

We investigated this question with {\it in-vitro} experiments, using the
platelets function analyzer Impact-R~\cite{shenkman:08}. The Impact-R
is a cylindrical device whose lower end is a fixed disk, serving as a
deposition surface, with a total area $S=132.7~mm^2$. It is covered
with polystyrene, on which platelets adhere and aggregate, but other
coating can easily be considered. The upper end of the Impact-R
cylinder is a rotating disk. Both disks are aligned with the
$xy$-plane. Anticoagulated (citrate) samples of whole blood were
loaded between these two disks, separated by $L=0.82~mm$. Due to the
rotation of the upper disk, the blood is subject to a laminar flow. A
controlled shear rate at the wall $\dot{\gamma}$ is created in a given
observation window of $1\times1~mm^2$, on the deposition surface.  We
imposed $\dot\gamma=100~s^{-1}$, corresponding to a value where the
platelet deposition reaches its maximum in a range of
$0-5000~s^{-1}$~\cite{supplementary}.

\begin{figure}
\includegraphics[width=.2\textwidth]{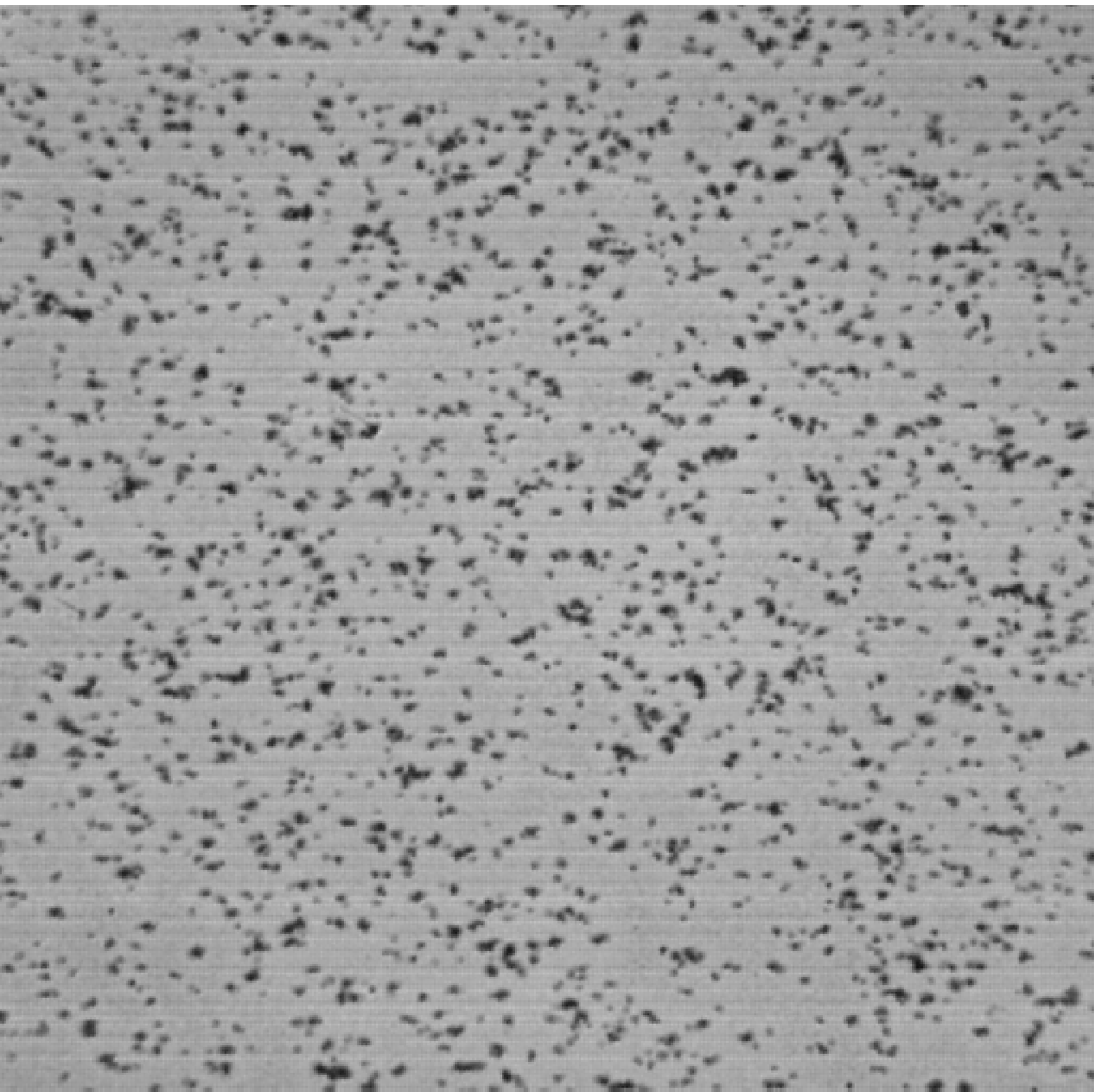}\hfill
\includegraphics[width=.2\textwidth]{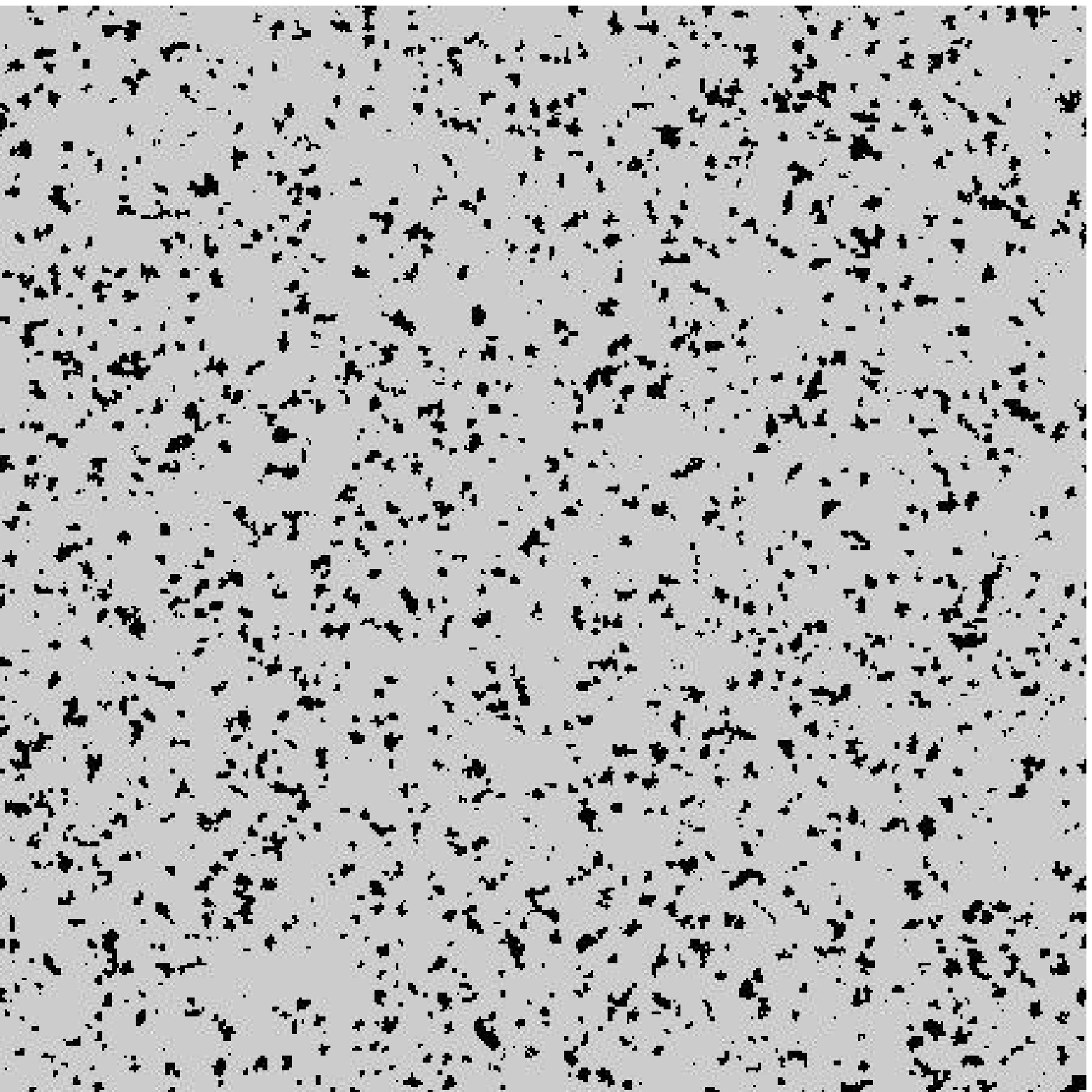}%
\caption{{\it Left:} platelet deposition as observed on the Impact-R
  $1mm\times1mm$ deposition window, after 300s. {\it Right:} Result of
  the numerical simulation, obtained with the parameters of
  table~\ref{table:simul}.}
\label{fig:impactR}.
\end{figure}

The experiment was repeated for 7 healthy donors. For each of them, we
measured on the observation window, the formation of clusters
resulting from deposition and aggregation of platelets, see
Fig.~\ref{fig:impactR}~left.
The number of clusters and their size, as well as the number of
activated platelets and non-activated platelets still in suspension were
measured at 20~$s$, 60~$s$, 120~$s$ and 300~$s$. For each of these
times a new experiment had to be performed, since the measurement
requires the interruption of the depositon process.

Our goal is to explain the observed time evolution of the above
quantities as a function of the parameters of the system. Numerous
mathematical models have been proposed in the literature to describe
the adhesion of platelets on a surface. See for
instance~\cite{tokarev:11,affeld:13} and reference therein. All the
models assume that platelets reach the deposition surface due to a
shear induced diffusion. The other parts of the proposed models depend
very much on the specific question addressed by the authors, and the
experimental device they have considered to produce their observations
(often assuming a steady state, which is not the case here). Our first
attempts to model the Impact-R experiments  convinced us that the
current knowledge is not sufficient to explain what is observed.
Therefore we developed an improved model that addresses the two most
unexpected features: (i) the well accepted Zydney-Colton model for
shear-induced diffusion is much too small to explain the observed
depletion rate of platelets from the suspension; (ii) there is a
mechanism that quickly limits the platelets deposition: during the
first $60~s$ there is a rapid increase of the number of clusters but,
at 300~$s$, there are still a lot of platelets in suspension but neither
formation of new clusters nor an increase of their surface.

\begin{figure}
\includegraphics[width=.45\textwidth]{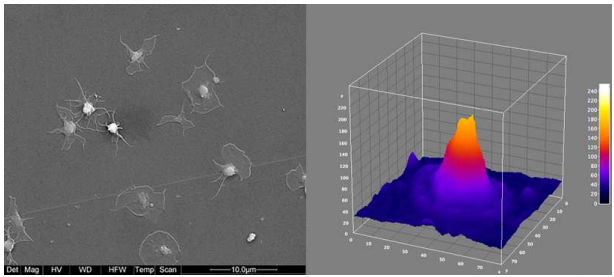}%
\caption{Left, Scan Electron Microscopy of platelet aggregates in the
  well. Right, the 3D shape of a platelets aggregate based on the
  optical height obtained by digital holographic microscopy (DHM).
  The vertical scale bar unit is $5.2nm$ and the field of view
  $12.8\mu m\times 12.8\mu m$.  }
\label{fig:3D}
\end{figure}

To explain the slowdown of the increase of the cluster areas, we first
assumed it might be due to a dominant growth in the 3rd
dimension ($z$-axis). Using a digital holographic microscope, we were able to
study the development of the aggregates thickness~\cite{frank:15}, a
quantity never measured before to our knowledge. See
fig.~\ref{fig:3D}.
However, the 3D nature of the aggregate was not sufficient to explain
the saturation of the areas of the clusters. We performed new
experiments~\cite{supplementary} that revealed that blood
albumin is the main factor limiting adhesion and aggregation, as it
competes with platelets for deposition.

To account for the above observation, we propose the following
scenario: (i) activated platelets (AP) adhere to the deposition surface, thus forming a seed
for a new cluster. (ii) non-activated platelets (NAP) and  AP can deposit at the
periphery, or on top, of an existing cluster. This is the aggregation
mechanism that makes the cluster grow. (iii) Albumin (Al) also
deposits on the surface, thus reducing locally the adhesion and
aggregation rates of platelets. 

The numerical model that we propose to describe and validate the above
process is the following.  The substrate $S$ is aligned with the
$xy$-plane and the blood layer expands along the $z$-axis. We consider
three types of particles: the AP, NAP, and Al. Any of them can be
either in suspension in the blood layer, or deposited on $S$. We
consider a mixed particle-density representation: particles in
suspension are described by a density field but as individual entities
when they deposit. The blood layer has thickness $L=0.82~mm$ and is
subject to a constant shear rate $\dot\gamma=100~s^{-1}$. This shear
induces a diffusion of the platelets and the albumin towards the
substrate.

The deposition substrate $S$ is discretized as $n\times n$ square
cells of area $\ds=5~(\mu m)^2$, corresponding to the size of a
deposited platelet (obtained as the smallest variation of cluster area
observed with the microscope). The number $n$ is chosen as $1000~\mu
m/\sqrt{5~(\mu m)^2}=447$ so that the total area of the substrate is
$S=1~mm\times1~mm$, as in the experiment.

Assuming a good horizontal mixing in the $xy-plane$ due to the
rotating flow, we consider that the densities of AP, NAP and Al only
vary along the $z$-axis and will be described by three 1D diffusion
systems: $\partial_t\rho=D\partial^2_z\rho$ where $\rho$ can be the
density of AP, NAP or Al. The initial condition is constant along $z$,
$\rho(z)=\rho_0$, determined by the experimental measurement, namely
172,200 $(\mu l)^{-1}$ for NAP, 4808 $(\mu l)^{-1}$ for AP, and
$2.69\times10^{13}~(\mu l)^{-1}$ for the Al. The boundary conditions
are (i) $J(L,t)=0$ at any time at the top of the blood layer, where
$J=-D\grad\rho$ is the flux of particles and $D$ the shear induced
diffusion. (ii) $\rho(0,t)\ds\dz=N(t)$ where $\dz$ is the thickness of
the boundary layer, {\it i.e.} the region above the substrate where
the particles are available for deposition. $N(t)$ is the average
number of particles in a volume $\ds\times\dz$ in this boundary
layer. The equation for $N$ reads
\begin{equation}
 \dot{N}=-J(0,t)\ds-p_{d}N(t)
\label{eq:N}
\end{equation}
where $p_{d}$ is the deposition
rate. The term $-J(0,t)$ corresponds to the new particles brought to
the boundary layer by the diffusion in the bulk.

For the adhesion-aggregation process, the particles are no longer
considered as densities, but as individual entities that can deposit
on any cell $(i,j)$ of the substrate. Let $\dt$ be the time
discretization and $p_d(i,j,t)$ the deposition rate on cell $(i,j)$ at
time $t$. For each cell,
$p_d(i,j,t)N(t)\dt$ is interpreted as the probability that a new
platelet (respectively new albumin) deposits on that cell. The time
step is chosen small enough so that this quantity is always smaller
than 1. The total number $m(t)$ of newly deposited particles on the entire
substrate is therefore
\begin{equation}
m(t)=\sum_{ij}[\mathrm{rand(i,j)}< p_d(i,j,t)N(t)\dt],
\end{equation}
where $\mathrm{rand(i,j)}$ are random numbers uniformly distributed in
$[0,1[$. The term $p_d N(t)$ in eq.~(\ref{eq:N}) is then computed as
$m/(n^2\dt)$, {\it i.e.} the average number of deposited particles per
cell and per time unit.

We explain below how the deposition rates $p_d(i,j,t)$ are implemented
in the model. If a cell is already occupied by a platelet, a new
platelet can only deposit on top, thus increasing the thickness of the
aggregate. This rate is noted $p_{top}$ and is constant over time. If
a cell is not occupied by a platelet, a new one can deposit. But the
deposition rate depends on the amount of albumin already in this
cell. It also differs for AP and NAP. The latter can only deposit in a
cell next to an existing aggregate. Al can only deposit in a cell
not occupied by a platelet.

More precisely, the deposition rules are the following.  An albumin
that reaches the substrate at time $t$ will deposit with a probability
$P(t)$ which depends on the local density $\rho_{al}(t)$ of already
deposited Al. We assume here that $P$ is proportional to the remaining
free space in the cell, $P(t)=p_{al}(\rho_{max}-\rho_{al}(t))$, where
$p_{al}$ is a parameter and $\rho_{max}$ is determined by the relative
size of an albumin and the size of a cell. We considered that 100,000
albumin particles can fit in the area $\ds$.

An activated platelet that hits a platelet-free cell on $S$ will
deposit with a probability $Q$, where $Q$ decreases as the local
concentration $\rho_{al}$ of albumin already present increases. We
assumed that $Q=p_{adh}\exp(-\lambda\rho_{al})$, where $p_{adh}$ and
$\lambda$ are parameters. This expression can be justified from the fact
that a platelet needs more free space than an albumin to attach to the
substrate, due to their size difference. As more albumin occupy the
deposition substrate, the probability to have enough space for the
platelet decreases roughly exponentially. This can be checked with a
simple deposition model on a grid, where small and large objects
compete for deposition.

Once an activated platelet has deposited, it is the seed of a new
cluster that will further grow due to the aggregation of other
platelets. In our model, NAP can deposit next to already deposited
platelets.  The aggregation probability $R$ is assumed to be
$R=p_{agg}\exp(-\lambda\rho_{al})$, with $p_{agg}$ another parameter.

The above model can be simulated numerically with different values of
the unknown parameters ($p_{al}$, $p_{adh}$, $\lambda$, $p_{agg}$
and $p_{top}$), in order to reproduce the time evolution of the
{\it in-vitro} experiment, namely the number $N_c(t)$ of clusters per
$mm^2$, the average cluster size $s(t)$ and the number $N_p(t)$ of
platelets still in suspension (activated and non-activated). 

\begin{figure}
\includegraphics[width=.5\textwidth]{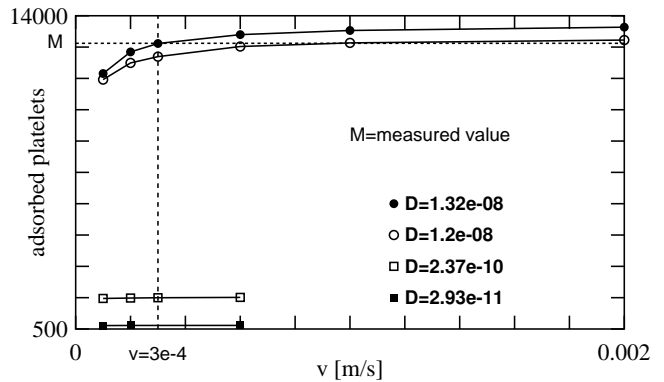}%
\caption{Number of particles adsorbed after 20~$s$, in the random walk
  model.}
\label{fig:adsorbed20s}
\end{figure}

However, the shear-induced diffusion coefficient $D$ and the thickness
of the boundary layer $\dz$ need to be determined. The value of $D$
from the Zydney-Colton theory (see
e.g~\cite{affeld:13,tokarev:11,aidun:15}) produces a flux of platelet
towards the deposition substrate $D\approx 5\times10^{-11}~m^2/s$ for
the direction perpendicular to the flow, when considering
$\dot\gamma=100~s^{-1}$ and a hematocrit of 40\%. This value is too
small to explain the 3125 platelets per $\mu\ell$ that have
disappeared on average from the bulk within the first 20~$s$ of the
experiment, even with a high deposition rate. In a column of section
$\ds=5~(\mu m)^2$ and height $L=0.82~mm$ ({\it i.e.} 4.1~ $\mu\ell)$,
this amounts to about 12800 deosited platelets. To determine $D$ we
take the smallest value compatible with the observed depletion of
platelet within the first $20~s$, assuming that all platelet hitting
the surface will adhere.  To solve this problem we use a
particle-based diffusion-deposition model. The experimentally
determined number of particle in a column of section $\ds=5~(\mu m)^2$
and height $L=0.82~mm$ are randomly distributed in space. Then they
are subject to a discrete time random walk: at every time step $\Dt$
each particle randomly changes its velocity to $\pm v$ and move
accordingly. Particles that cross the line $z=0$ are removed. Those
reaching $z=L$ are bounced back from were they came.  Using that
$D=v^2\Dt/2$, we can chose $\Dt$ as a function of $D$ and explore the
deposition count after $20~s$ for different $v$ and $D$, see
Fig.~\ref{fig:adsorbed20s}.
To match the experimental observation, assuming a perfect adsorption,
the diffusion coefficient $D$ needs to be $D\ge 1.2\times10^{-8}$. The published
value $D\approx5\times10^{-11}~m^2/s$ for $\dot{\gamma}=100~s^{-1}$ and
hematocrit of 40\% is clearly too small to explained the observed
deposition rate, whatever $v$ is.
In what follows we chose $D=1.3~10^{-8}~m^2/s$, corresponding to the molecular 
velocity $v=3\times 10^{-4}~m/s$, as obtained
from a fully resolved simulation of platelets and red blood cells in a
shear rate of $100~s^{-1}$ and hematocrit of $40~\%$~\cite{uva}.
This discrete time random-walk model can also be used to determine
$\dz$, the thickness of boundary layer that is needed for a continuous
description of the diffusion-deposition process. Numerical
investigations show that $\dz=2~10^{-5}~m$ produces a good agreement
between the continuous and discrete models.

The full model for adhesion and aggregation of platelets, including
the presence of albumin, has been run for 5 min of physical time, with
$D=1.3~10^{-8}~m^2/s$ and $\dz=2~10^{-5}~m$. We checked the
independence of the result with respect to spatial and temporal
discretization of the 1D diffusion models, which is not obeyed
if the boundary layer $\dz$ is omitted.

\begin{figure}
\includegraphics[width=.5\textwidth]{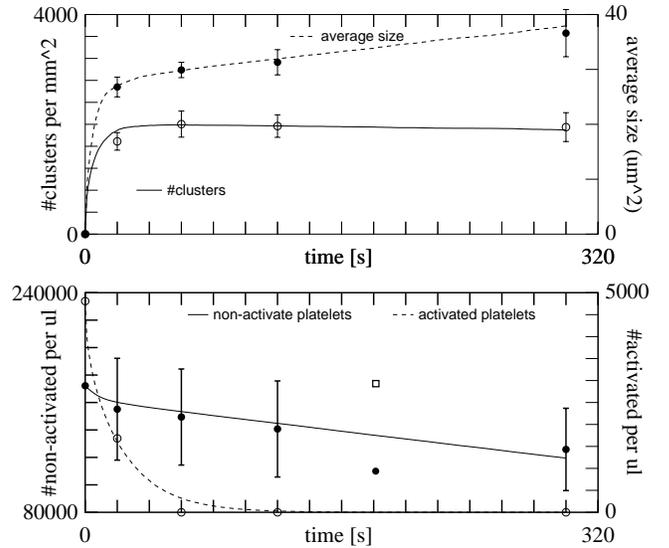}%
\caption{The result of the adhesion and aggregation model (continuous
  and dashed lines) and the experimental data (points).}
\label{fig:deposition-results}
\end{figure}

The value of $p_{al}$, $p_{adh}$, $\lambda$, $p_{agg}$ and $p_{top}$
that give the best agreement with experimental observation have been
obtained by exploring the parameter space (see
table~\ref{table:simul}.  The high value of $p_{adh}$ suggests that
the adhesion is diffusion-limited, whereas aggregation is
reaction-limited.  Figure~\ref{fig:impactR}~(right) shows the
simulated deposition pattern after 5 min.
Fig.~\ref{fig:deposition-results} shows that our model reproduces
quantitatively the {\it in-vitro} measurements, thus confirming the
proposed scenario of deposition and aggregation, and the competition
with albumin.  Note that the values in Table~\ref{table:simul} are
expected to depend on the shear rate $\dot{\gamma}$ and the nature of
the substrate.

\begin{figure}
\includegraphics[width=.24\textwidth]{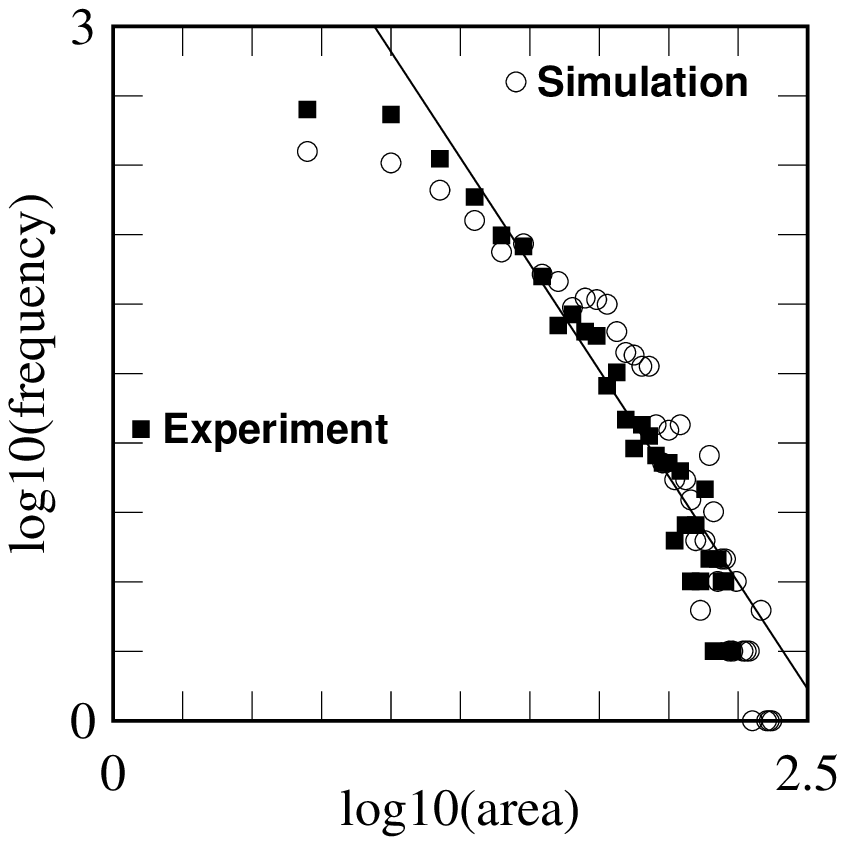}\hfill
\includegraphics[width=.24\textwidth]{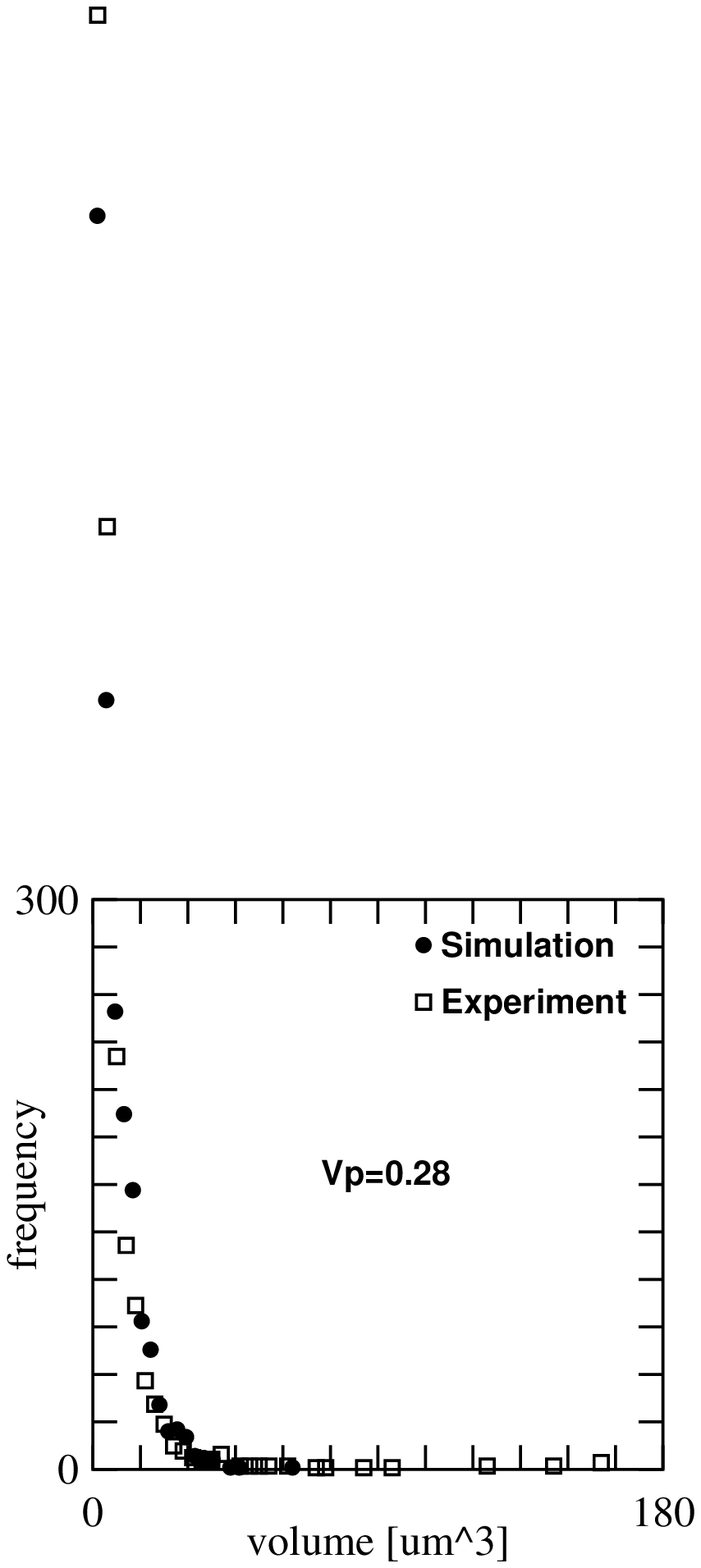}%
\caption{{\it Left:} Distribution of the size of clusters at
  $t=300~s$. The area is given in $(\mu m)^2$. {\it Right:}
  Distribution of cluster volumes at $t=300~s$. The data suggest a
  power law with exponent -2.8. Simulation parameters are given in
  table~\ref{table:simul}. }
\label{fig:distribution}
\end{figure}

To further test the model against the experiment, we compared the
distribution of sizes $s$ and volumes of the clusters measured by
digital holographic microscopy (see Fig. \ref{fig:distribution}).  The
size distribution from the simulation matches very well that of the
experiment, at $20~s$ (data not shown) and at $300~s$. The log-log
plot suggests a power law distribution $p(s,t)=\alpha(t)
s^{-\beta(t)}$, with $\beta(20s)=2.23$ and $\beta(300s)=1.84$. The
simulation gives the distribution of the number of particles per
clusters, including those that piled up along the $z$-axis. To compare
this distribution with the experimental one, one needs to assign a
volume to each platelet.  Observations with the holographic
microscope~\cite{frank:15} indicate that the average volume of a
platelet in suspension ($6~(\mu m)^3$) is too large since the
platelets lose a significant part of its volume while
depositing. Using a volume per platelet of $V_{p}=0.28~(\mu m)^3$ with
the simulation data makes the volume distribution compatible with the
{\it in-vitro} measurements (see Fig.~\ref{fig:distribution}~(right)). This
value of $V_{p}$ is compatible with the meaurement of the total volume
of the deposited aggregates, found experimentally to be
$V_{tot}=1'701272~(\mu m)^3$, and the total number of deposited
platelets, $N_{tot}=6,032,000$.

\begin{table}
\vskip .2cm
\begin{tabular}{|c|c|c|c|c|}
\hline
$p_{adh}~[s^{-1}]$ & $p_{agg}~[s^{-1}]$ & $p_{al}~[s^{-1}]$ &$p_{top}~[s^{-1}]$ & $\lambda~[(\mu m)^2]$
\\ \hline
110        & 14.6     & $1.7\times10^{-3}$ &0.6   & 30 
\\
\hline
\end{tabular}
\caption{The parameters of the model found to provide the best 
  fit of the experimental data (see Fig.~\ref{fig:deposition-results}).}
\label{table:simul}.
\end{table}

This study has revealed new phenomena in the adhesion-aggregation
processes of platelets, namely the competition with albumin and the
different roles played by activated and non-activated platelets. By
adhesion, activated platelets initiate a new cluster which mostly
grows due to the non-activated platelets. We have also analyzed the 3D
structures of the aggregates and obtained the distribution of areas
and volumes. By tuning the model parameters so as to fit the {\it
  in-vitro} time observations, the adhesion and aggregation rates can
be measured.  The excellent agreement between the model and the
experiment gives a strong credit to the plausibility of the scenario
we have considered. Our study has also revealed that the Zydney-Colton
shear induced diffusion coefficient is significantly too small to
explain the observed deposition rate. It seems that other researchers
have already experienced a similar problem~\cite{aidun}.

We acknowledge support from the European FP7-VPH THROMBUS project and
from CADMOS.

%

\end{document}